JP6.12  MICROBURST WINDSPEED POTENTIAL ASSESSMENT:
PROGRESS AND DEVELOPMENTS

Kenneth L. Pryor
Center for Satellite Applications and Research (NOAA/NESDIS)
Camp Springs, MD

## 1. INTRODUCTION

A suite of products has been developed and evaluated to assess hazards presented by convective downbursts to aircraft in flight derived from the current generation of Geostationary Operational Environmental Satellite (GOES) (8-P). The existing suite of GOES microburst products employs the GOES sounder to calculate risk based on conceptual models of favorable environmental profiles for convective downburst generation. Large output values of the microburst index algorithms indicate that the ambient thermodynamic structure of the troposphere fits the prototypical environment for each respective microburst type (i.e. Wet, Hybrid, Dry, etc.). Accordingly, a diagnostic nowcasting product, the Microburst Windspeed Potential Index (MWPI) (Pryor 2008a), is designed to infer attributes of a favorable microburst environment: a convective boundary layer with a steep temperature lapse rate and low relative humidity in the surface layer. These conditions foster intense convective downdrafts due to evaporational cooling as precipitation descends in the sub-cloud layer and results in the generation of negative buoyancy. A concern pertaining to the GOES sounder products is the current temporal and spatial resolution (60 minutes, 10 km). The GOES-R Advanced Baseline Imager (ABI) has promising capability as a sounder with greatly improved temporal and spatial resolution as compared to the existing GOES (8-P) sounders. However, until GOES-R soundings and associated derived products are operational, a need has been established for a higher resolution GOES-derived microburst risk product.

*Corresponding author address*: Kenneth L. Pryor, NOAA/NESDIS/STAR, 5200 Auth Rd., Camp Springs, MD 20746-4304; e-mail: Ken.Pryor@noaa.gov.

A multispectral GOES imager product (Pryor 2008b) has been developed and experimentally implemented to assess downburst potential over the western United States with improved temporal and spatial resolution. The availability of the split-window channel in the GOES-11 imager allows for the inference of boundary layer moisture content. Wakimoto (1985), based on his study of microbursts that occurred during the Joint Airport Weather Studies (JAWS) project, noted favorable environmental conditions over the western United States: (1) intense solar heating of the surface and a resulting superadiabatic surface layer; (2) a deep, dry-adiabatic convective boundary layer (Sorbjan 1989) that extends upward to near the 500mb level; (3) a well-mixed moisture profile with a large relative humidity gradient between the mid-troposphere and the surface. The GOES-West (GOES-11) imager microburst algorithm employs brightness temperature differences (BTD) between band 3 (upper level water vapor, 6.7µm), band 4 (longwave infrared window, 10.7µm), and split window band 5 (12µm). Band 3 is intended to indicate mid to upper-level moisture content and advection while band 5 indicates low-level moisture content. Soden and Bretherton (1996) (SB96), in their study of the relationship of water vapor radiance and layer-average relative humidity, found a strong negative correlation between 6.7µm brightness temperature ($T_b$) and layer-averaged relative humidity (RH) between the 200 and 500-mb levels. Thus, in the middle to upper troposphere, decreases in $T_b$ are associated with increases in RH as illustrated in Figure 4 of SB96. It follows that large BTD between bands 3 and 5 imply a large relative humidity gradient between the mid-troposphere and the surface, a condition favorable for strong convective downdraft generation due to evaporational cooling in the sub-cloud layer. In addition, small BTD between bands 4 and 5 indicate a relatively dry surface layer with solar heating in progress. Thus the GOES imager microburst risk (MBR) product is based on the following algorithm in which the output

brightness temperature difference (BTD) is proportional to microburst potential:

$$MBR (BTD) = \{T_5 - T_3\} - \{T_4 - T_5\} \quad (1)$$

Where the parameter $T_n$ represents the brightness temperature observed in a particular imager band. The relationship between BTD and microburst risk in the product image is based on the following assumptions: (1) A deep, well-mixed convective boundary layer exists in the region of interest; (2) moisture for convective storm development is based in the mid-troposphere and is advected over the region of interest; and (3) the mid- to upper-tropospheric layer of moisture is vertically extensive and would yield precipitation if provided a sufficient forcing mechanism.

This product provides a higher spatial (4 km) and temporal (30 minutes) resolution than is currently offered by the GOES sounder microburst products (10 km, 60 minutes) and thus, provides useful information to supplement the sounder products in the convective storm nowcasting process. In addition, this imager product provides microburst risk guidance in high latitude regions, especially north of latitude 50°N, where existing sounder coverage is not available.

Both the GOES sounder MWPI and imager microburst risk products are predictive linear models developed in the manner exemplified in Caracena and Flueck (1988). Each microburst product consists of a set of predictor variables that generates output of expected microburst risk. This paper compares and contrasts the sounder and imager microburst products and outlines the advantages of each product in the nowcasting process. An updated assessment of the sounder MWPI and imager microburst products, case studies demonstrating effective operational use of the microburst products, and validation results for the 2008 convective season over United States Great Plains and Great Basin regions will be presented.

## 2. METHODOLOGY

### 2.1 *GOES Imager Microburst Product*

The objective of this validation effort was to qualitatively and quantitatively assess the performance of the GOES imager-derived microburst product by employing classical statistical analysis of real-time data. Accordingly, this effort entailed a study of downburst events over the Eastern Snake River Plain (ESRP) of southeastern Idaho during the convective seasons of 2007 and 2008 that was executed in a manner that emulates historic field projects such as the 1982 Joint Airport Weather Studies (JAWS) (Wakimoto 1985). GOES-11 image data was collected for pre-convective environments associated with 25 downburst events that occurred within the Idaho National Laboratory (INL) mesonet domain during July and August 2007 and between June and September 2008. Clawson et al. (1989) provides a detailed climatology and physiographic description of the INL as well as a description of the associated mesonetwork. Wakimoto (1985) and Atkins and Wakimoto (1991) discussed the effectiveness of using mesonetwork surface observations and radar reflectivity data in the verification of the occurrence of downbursts. Well-defined peaks in wind speed (Wakimoto 1985; Atkins and Wakimoto 1991) were effective indicators of downburst occurrence. It was found that derived product imagery generated one to three hours prior to downburst events provided an optimal characterization of the pre-convective thermodynamic environment over the INL.

During the 2007 convective season, product images were generated by Man computer Interactive Data Access System (McIDAS)-X where GOES imager data was read and processed, brightness temperature differences were calculated, and risk value contours were overlain on GOES imagery. The image data consisted of derived brightness temperatures from infrared bands 3, 4, and 5, obtained from the Comprehensive Large Array-data Stewardship System (CLASS, http://www.class.ncdc.noaa.gov/) for archived data and the McIDAS Abstract Data Distribution Environment (ADDE) server for real-time data. During the 2008 convective season, microburst algorithm output was also visualized by McIDAS-V software (version 1.0alpha10) (Available online at http://www.ssec.wisc.edu/mcidas/software/v/).
McIDAS-V is a no-cost, open source, visualization and data analysis software package that displays weather satellite and other geophysical data in two- and three-dimensions. McIDAS-V can also analyze and manipulate satellite data with powerful mathematical functions. McIDAS-V was used to visualize output area files that were generated by McIDAS-X from archived data. A contrast stretch and built-in color enhancement was applied to the output images to highlight regions of elevated microburst risk. In addition, the three-band microburst risk algorithm was implemented in McIDAS-V, where, for selected events, the program was utilized to retrieve real-time GOES imager data and produce a color-enhanced product image. Visualizing algorithm output in McIDAS-V allowed for cursor interrogation of output brightness temperature and more precise recording of BTD values associated with observed downburst events. This methodology should serve to increase the statistical significance of the relationship between output BTD and microburst wind gust magnitude.

Downburst wind gusts, as recorded by National Oceanic and Atmospheric Administration (NOAA) mesonetwork observation stations, were measured at a height of 15 meters (50 feet) above ground level. Archived NOAA mesonet observations are available via the NOAA INL Weather Center website (http://niwc.noaa.inel.gov). In order to assess the predictive value of the GOES imager microburst product, the images used in validation were obtained for valid times one to three hours prior to the observed surface wind gusts. Derived images generated from the infrared dataset consisted of the following: (1) band 5-3 difference image; (2) band 4-5 difference image; (3) microburst risk image and (4) color-enhanced microburst risk image derived from the algorithm as defined in equation 1. In the derived images, a contrast stretch and contouring of output brightness values were employed to highlight regions of elevated microburst risk. In addition, based on the statistical relationship derived between output BTD and observed downburst wind gust speed, a color-enhanced risk image was generated that indicates increasing microburst potential as a progression from yellow to red shading. Transmittance weighting functions from bands 3, 4, and 5 that specify the relative contribution from each atmospheric layer to emitted radiation were obtained over Boise, Idaho (closest radiosonde observation (RAOB) site to INL) from a Cooperative Institute for Meteorological Satellite Studies (CIMSS) website (http://cimss.ssec.wisc.edu/goes/wf/index.php). The weighting functions were compared to temperature and moisture profiles in the 0000 UTC Boise RAOB and product imagery for selected microburst events that occurred during the evening hours (after 0000 UTC or 1700 LST). Thus, the group of spectral bands selected to detect radiation emitted from layers of interest in the atmosphere could be determined to be most effective for temperature and moisture profiling for the purpose of inferring the presence of a favorable environment for microbursts.

For each microburst event, product images were compared to radar reflectivity imagery and surface observations of convective wind gusts as provided by INL mesonet stations. Next Generation Radar (NEXRAD) base reflectivity imagery (levels II and III) from National Climatic Data Center (NCDC), University Corporation for Atmospheric Research (UCAR) Internet Data Distribution (IDD) server, and Iowa Environmental Mesonet (IEM) was utilized to verify that observed wind gusts were associated with downbursts and not associated with other types of convective wind phenomena (i.e. gust fronts). Another application of the NEXRAD imagery was to infer microscale physical properties of downburst-producing convective storms. Particular radar reflectivity signatures, such as the rear-inflow notch (RIN)(Przybylinski 1995) and the spearhead echo (Fujita and Byers 1977), were effective indicators of the occurrence of downbursts. In addition, radar temperature profiles over the INL, generated at Grid3 mesonet station, were archived as a means to validate quantitative information as portrayed in the derived product images.

Covariance between the variables of interest, microburst risk (expressed as output brightness temperature) and surface downburst wind gust speed, was analyzed to assess the performance of the imager microburst algorithm. A very effective means to quantify the functional relationship between microburst index algorithm output and downburst wind gust strength at the surface was to calculate correlation between these variables. Thus, correlation between GOES imager microburst risk and observed surface wind gust velocities for the selected events were computed to assess the significance of this functional relationship. Statistical significance testing was conducted, in the manner described in Pryor and Ellrod (2004), to determine the confidence level of correlations between observed downburst wind gust magnitude and microburst risk values. The derived confidence level is intended to quantify the robustness of the correlation between microburst risk values and wind gust magnitude.

## 2.2 *GOES Sounder Microburst Product*

In a similar manner to the imager product, this validation effort assessed and intercompared the performance of the GOES sounder-derived microburst products by employing classical statistical analysis of real-time data. Accordingly, this effort entailed a study of downburst events over two different microburst environments: the southern High Plains and northern Great Basin regions. Again, this validation was executed in a manner that emulated historic field projects such as the 1982 Joint Airport Weather Studies (JAWS) (Wakimoto 1985) and the 1986 Microburst and Severe Thunderstorm (MIST) project (Atkins and Wakimoto 1991). Data from the GOES MWPI product was collected over the Oklahoma Panhandle and western Texas for downburst events that occurred between 1 June and 30 September 2007 and over Oklahoma, western Texas and southeastern Idaho for events that occurred between 1 June and 31 August 2008. Microburst index values were then validated against surface observations of convective wind gusts as recorded by Oklahoma, West Texas, and INL Mesonet stations. Wakimoto (1985) and Atkins and Wakimoto (1991) discussed the effectiveness of using mesonet surface observations and radar reflectivity data in the verification of the occurrence of downbursts. Well-defined peaks in wind speed as well as significant temperature decreases (Wakimoto 1985; Atkins and Wakimoto 1991) were effective indicators of high-reflectivity downburst occurrence. As illustrated in the flowchart in Figure 11 of Pryor (2008a), images were generated in Man computer Interactive Data Access System (McIDAS) by a program that reads and processes GOES sounder data, calculates and collates microburst risk values, and overlays risk values on GOES imagery. Output images were then archived on the STAR FTP server.

Since surface data quality is paramount in an effective validation program, Oklahoma, western Texas, and southeastern Idaho were chosen as a study regions due to the wealth of high quality surface observation data provided by the Oklahoma, West Texas, and INL Mesonets (Brock et al. 1995; Schroeder et al. 2005; Clawson et al. 1989), a thermodynamic environment typical of the High Plains and Great Basin regions during the warm season, and relatively homogeneous topography. Pryor (2008a) discussed the importance of the dryline (Schafer 1986) in convective storm climatology in the Southern Plains region as well as the selection of the High Plains region for a validation study.

Downburst wind gusts, as recorded by mesonet observation stations, were measured at a height of either 10 meters (33 feet) or 15 meters (50 feet) above ground level. In order to assess the predictive value of GOES microburst products, the closest representative index values used in validation were obtained for retrieval times one to three hours prior to the observed surface wind gusts. Representativeness of proximate index values was ensured by determining from analysis of surface observations, and radar and satellite imagery, that no change in environmental static stability and air mass characteristics between product valid time and time of observed downbursts had occurred. Furthermore, in order for the downburst observation to be included in the validation data set, it was required that the parent convective storm cell of each downburst is located in proximity. Previous field studies of microbursts, especially the MIST project in northern Alabama (Atkins and Wakimoto 1991), noted that peak wind gusts are typically recorded three to five kilometers from the point of impact. Beyond a distance of five kilometers, interaction of the convective storm outflow with the surface, most likely through the process of friction, will result in a decrease in downburst wind velocity and subsequently, wind gust measurements that are unrepresentative of the ambient thermodynamic environment. Ideally, at the time of downburst occurrence, the distance between the parent storm cell and the location of the measured downburst wind gust should be no more than five kilometers. An additional criterion for inclusion into the data set was a wind gust measurement of at least F0 intensity (35 knots) on the Fujita scale (Fujita 1971). Wind gusts of 35 knots or greater are considered to be operationally significant for transportation, especially boating and aviation. An algorithm devised by Wakimoto (1985) to visually inspect wind speed observations over the time intervals encompassing candidate downburst events was implemented to exclude gust front events from the validation data set. In summary, the screening process employed to build the validation data set that consisted of criteria based on surface weather observations and radar reflectivity data yielded a sample size of 85 downbursts and associated index values.

Similar to the validation process for the imager microburst product, for each microburst event, product images were compared to radar reflectivity imagery. Next Generation Radar (NEXRAD) base reflectivity imagery from the Texas Tech University archive (http://wedgefest.wind.ttu.edu/rad_images/) was utilized for this purpose.

Covariance between the variables of interest, MWPI and surface downburst wind gust speed, was considered. Algorithm effectiveness was assessed as the correlation between MWPI values and observed surface wind gust velocities. Statistical significance testing was conducted to determine the confidence level of the correlation between observed downburst wind gust magnitude and microburst risk values.

## 3. CASE STUDIES

### 3.1 *Idaho Microbursts*

During the afternoon of 30 and 31 August 2008, respectively, areas of convective storms developed over the mountains of extreme southern Idaho and northeastern Nevada and tracked northeastward through the Idaho National Laboratory (INL) complex. These events were of great interest due to the simultaneous availability of both GOES MWPI and imager microburst risk products two to three hours prior to the observed downbursts over the INL complex. Both product images effectively characterized the pre-downburst environment. This case also proved to be efficacious in the comparison of different displays of the imager microburst product.

Figure 1 displays a GOES-11 imager derived microburst risk product at 2330 UTC and the corresponding sounder MWPI image at 0000 UTC 31 August, respectively. Apparent in the GOES-11 imager product is convective storm activity, displayed as cellular or globular dark regions, developing over extreme southern Idaho. Forcing for this convective storm activity was most likely solar heating of the mountain ridges in the presence of significant mid-level moisture that was advected from the tropical North Pacific Ocean. GOES-11 Northern Hemisphere water vapor imagery (not shown) indicated that the deep convective storm activity developed within the plume of moisture that originated over the tropical Pacific Ocean. The convective storm activity propagated into the Snake River Plain where intense surface heating during the afternoon hours resulted in the development of a deep, dry-adiabatic boundary layer. As shown in figure 1, elevated output BTD (>40°K, light grey shading) near the location of downburst occurrence (46 knots, plotted in image) substantiated the presence of a well-developed mixed layer. As discussed earlier, large BTD between bands 3 and 5 implies a large relative humidity gradient and well-mixed moisture profile between the mid-troposphere and the surface, a condition favorable for strong convective downdraft generation due to evaporational cooling as precipitation descends in the deep sub-cloud layer. The corresponding MWPI image displayed the closest representative index value of 25 near Minidoka (location "M") mesonet station that indicated wind gust potential of 35 to 49 knots, based on statistical relationships developed for downbursts that occur over the southern High Plains. Considering the microburst wind gust intensity of 46 knots recorded at Radioactive Waste Management (RWM) mesonet station at 0215 UTC, it is evident that the MWPI product echoed optimal boundary layer conditions for microbursts.

Figure 2 displays GOES-11 imager transmittance weighting functions over Boise (location "B" in product image). The black curve represents the weighting function for band 3 (6.7µm) while the tan and brown curves represent functions for bands 4 (11µm) and 5 (12 µm), respectively. The band 3 weighting function peak was located near the 500-mb level, corresponding roughly to the mid-tropospheric moist layer displayed in the RAOB. The band 3 weighting function peak, overlying peaks in bands 4 and 5 functions at the surface, correlated well with the classic "inverted-v" sounding profile. The combined display of the RAOB and weighting functions indicated that the mid-tropospheric moist layer was vertically extensive and was overlying a dry, convective boundary layer, thus, illustrating favorable environmental conditions for strong convective downdrafts. Microburst risk was calculated over Boise using brightness temperatures associated with the GOES-11 weighting functions. A BTD of 51 was associated with the RAOB profile over Boise at 0000 UTC, very close to the value (46) associated with the 46-knot downburst wind gust recorded at Radioactive Waste Management (RWM) mesonet station at 0215 UTC. Note similar shading near Boise and RWM station

(orange) in the color-enhanced risk image that indicated an elevated risk of microbursts.

Surface observations as displayed in the meteograms in Figure 3 confirm the presence of favorable boundary layer conditions for microbursts during the evening hours of 30 August. Especially noteworthy is a large surface dewpoint depression near 32°C and relative humidity less than 20% during the two hours prior to downburst occurrence at RWM station. In addition, temperature lapse rates in the surface layer were determined to be superadiabatic (>10°C/km) and nearly dry-adiabatic (9 °C/km) in the boundary layer based on comparisons between temperatures measured at 2 meters and 15 meters above ground level at RWM and the radar temperature profile generated at Grid3 (not shown) about two hours prior to downburst occurrence. Superadiabatic lapse rates are often associated with strong insolation and resulting surface heating. Accordingly, downburst occurrence was apparent in the mesonet meteogram, characterized by a sharp peak in wind speed and abrupt change in wind direction.

Composite base reflectivity imagery displayed that the low-reflectivity, downburst-producing convective storm developed over southern Butte County between 0145 and 0200 UTC and tracked northeastward toward the INL complex. By 0210 UTC, the cell had evolved into a spearhead echo configuration over the southwestern portion of the INL complex as displayed in the radar image in Figure 4. The image indicates that the location of peak downburst winds ("R"), as expected, occurred near the apex of the spearhead echo five minutes later. The storm cell continued to propagate northeastward and dissipated over central Butte County by 0230 UTC. Low radar reflectivity (blue shading, <20 dBZ) in the presence of large subcloud lapse rates (9 °C/km) strongly suggests that the microburst was of the "dry" type.

During the following afternoon of 31 August 2008, stronger downbursts were observed over the INL mesonet domain. Measured convective wind gusts between 50 and 55 knots were associated with significantly higher output BTD and elevated MWPI values. Similar to the 30 August event, downbursts occurred shortly after a maximum in surface heating when the boundary layer was dry and sufficiently well-mixed. However, convective storm activity that developed during the afternoon of 31 August was fueled by a plume of mid-level moisture that originated over the central North Pacific Ocean. Downburst activity occurred on the northwestern periphery of the area of convective storms where the surface layer was especially dry with relative humidity below 20%. As shown in Figure 5, convective storm activity developed over the mountains of northeastern Nevada and extreme southern Idaho. Convective storms propagated northeastward over the eastern portion of the INL mesonet domain between 2000 and 2200 UTC. Peak downburst wind gusts of 52 and 55 knots were recorded Experimental Breeder Reactor (EBR, location "E" in Figure 5) and Rover (ROV, just north of location "E" in Figure 5) INL mesonet stations, respectively, between 2125 and 2135 UTC. It is apparent in Figure 5 that the severe downbursts occurred in close proximity to local maxima in GOES imager BTD values (near 60K) and sounder MWPI values (19 to 26). The EBR meteogram in Figure 3 provides evidence of favorable boundary layer conditions for microbursts at 1900 UTC (1200 MST) with a large surface dewpoint depression (22°C) and low relative humidity (near 20%). Composite base reflectivity imagery in Figure 6, near the time of downburst occurrence at EBR, revealed that convective winds were again associated with spearhead echo, however maximum reflectivity in the parent storm was considerably higher (45 to 50 dBZ, orange shading). The combination of a superadiabatic temperature lapse rate in the surface layer at EBR and high radar reflectivity associated with the parent cell suggested the downburst was of the "hybrid" type, driven by a combination of precipitation loading and sub-cloud evaporational cooling. The more hybrid nature of the microburst was apparent in the EBR meteogram with a marked temperature decrease (3C° in 15 minutes) at the surface between 2130 and 2145 UTC. Thus, the microburst events of 30 and 31 August over the INL complex demonstrate the ability of both the MWPI and imager products to effectively indicate potential through a significant portion of the microburst environment spectrum.

### 3.2 *West Texas Downbursts*

During the afternoon of 27 July 2008, near 2300 UTC, a strong downburst, with an associated wind gust of 63 knots, occurred over western Texas in close proximity to elevated GOES MWPI values. The GOES MWPI product image indicated wind gust potential of 50 to 64 knots at 2100 UTC near Turkey (West Texas) mesonet station where the severe downburst was recorded two hours later. The ambient thermodynamic environment over western Texas was typical of the southern Great Plains with strong static instability and a deep, well-mixed boundary layer that favored the development of intense convective downdrafts and resultant downburst generation. Strong downbursts also occurred elsewhere in the Oklahoma and Texas Panhandles.

Apparent in Figure 7 is convective storm activity that developed over western Texas and the Oklahoma Panhandle between 2000 and 2300 UTC 27 July. A peak downburst wind gust of 63 knots (plotted in images) was recorded at Turkey (West Texas) mesonet station near 2300 UTC. Note that the downburst occurred in close proximity to one of the higher MWPI values over western Texas (68), and in proximity to elevated GOES imager microburst risk values. The sounding in Figure 8 displayed a shallow "inverted-v" profile typical of the southern Great Plains during the convective season with large CAPE, a well-mixed boundary layer, and a mid-tropospheric dry air layer that favored the development of intense convective downdrafts. Based on linear regression, as illustrated in Figure 10, an MWPI value of 68 is correlated with wind gust potential of 55 knots, close to the observed wind gust speed of 63 knots at Turkey.

A meteogram from Turkey mesonet station, also in Figure 8, indicates that downburst occurrence was apparent as a sharp peak in wind speed and abrupt change in wind direction near 2300 UTC (1700 CST), similar to the Idaho downbursts displayed in Figure 3. Also similar to the pre-downburst environment described in the Idaho downburst case, was a large surface dewpoint depression (40°F/22°C) and a superadiabatic temperature lapse rate about two hours prior to the occurrence of severe convective winds. However, a major difference between the Idaho and West Texas downbursts was the significant temperature drop that was observed at Turkey mesonet station. Sharp temperature decreases are often associated with wet microbursts. Another major difference between the West Texas and Idaho downburst events by comparing Figures 3 and 8 is higher surface dewpoints over western Texas that characterized the pre-convective environment illustrated in Figure 7.

In this case, higher surface dewpoints and large CAPE translated into heavy precipitation within convective storms that served as an initiating mechanism for downbursts. The larger storm precipitation content was apparent in the NEXRAD image in Figure 9. The isolated parent convective storm that produced the Turkey downburst exhibited high reflectivity (>50 dBZ) and a bow echo structure (Przybylinski 1995) in which the apex of the bow was located in close proximity to Turkey station (XTUR) at the time of downburst occurrence. A weaker downburst (46 knots) occurred about 30 minutes later at Spur (XSPR) mesonet station, where a lower MWPI value (47) was indicated at 2300 UTC (not shown). The combination of high radar reflectivity and a superadiabatic lapse rate in the surface layer suggests that hybrid microbursts occurred during the afternoon of 27 July over western Texas. In a similar manner to the dry and hybrid microburst events over Idaho, the MWPI product effectively indicated the potential for severe wind gusts of 50 to 64 knots associated with the microbursts that occurred over western Texas.

More recently, during the afternoon of 8 December 2008, strong downbursts occurred over eastern New Mexico and western Texas. Clusters of convective storms developed over the region ahead of an upper-level cyclone. Strong downbursts, produced by a convective storm line that developed over eastern New Mexico between 2000 and 2100 UTC, occurred in close proximity to elevated GOES sounder MWPI and GOES imager microburst risk values, as indicated in Figure 11. Downburst wind gusts in excess of 50 knots were recorded by West Texas Mesonet stations near the New Mexico border between 2055 and 2115 UTC as the storm line tracked east. The ambient thermodynamic environment over western Texas during the afternoon of 8 December was more typical of the southern Great Plains warm season with a deep, well-mixed boundary layer that favored the development of intense convective downdrafts and resultant downburst generation. The downbursts resulted in the generation of a dust storm over western Texas that affected the Lubbock area (http://www.mesonet.ttu.edu/cases/DuststormI_120808/Dec-08-2008Duststorm.html). Apparent in the product images in Figure 11 are clusters of convective storms over the western Texas Panhandle and over eastern New Mexico that would track eastward over western Texas during the following three hours. Associated with the convective storm cluster near the New Mexico border were downburst wind gusts of 50 and 57 knots (plotted in images) that were recorded by

Plains and Denver City (West Texas) mesonet stations at 2115 UTC. Note that the downbursts occurred in close proximity to elevated MWPI and imager microburst risk values. Thus, the sounder and imager microburst products effectively indicated the potential for strong convective wind gusts.

## 4. STATISTICAL ANALYSIS AND DISCUSSION

Analysis of covariance between the variables of interest, microburst risk (expressed as output brightness temperature) and surface downburst wind gust speed, provided favorable results for the imager microburst product. Favorable results include a strong correlation ($r=.68$) and a small mean difference (1.68) between microburst risk values and wind gust speed (in knots) for a dataset of 25 microburst events that occurred during the 2007 and 2008 convective seasons. In addition, statistical significance testing revealed a high (91%) confidence level that the correlation did represent a physical relationship between risk values and downburst magnitude and was not an artifact of the sampling process. The scatterplot in Figure 10 of BTD values vs. microburst wind gust speed displays this strong correlation and the linear predictive model ($y=.7641x+12.646$) where the variable $x$ represents a set of predictors and the variable $y$ represents the response variable.

Also shown in Figure 10 is the scatterplot of MWPI values vs. measured wind speeds that effectively illustrates the strong correlation ($r=.74$, $r^2=.54$) between MWPI values and wind gust magnitude, in which 54% of the variability in wind gust speed is coupled with variability in MWPI. As illustrated in the scatterplot diagram, MWPI values do not represent absolute wind gust speeds, but rather indicate relative convective wind gust potential that can be statistically related (or correlated) to surface wind gust speed. Similar to the imager microburst product, the MWPI product is a linear predictive model ($y=.3571x+31.436$). The statistical correlation approach allows the user to tune the relationship between MWPI values and downburst wind gust speeds according to local climatology. One important goal of this validation effort was to derive a relationship between MWPI values and categories of microburst wind gust potential (i.e. <35 knots, 35-49 knots, 50-64 knots, etc.) utilizing the linear predictive model. Diurnal and inter-diurnal trends in microburst activity found in this study echo the findings documented in Pryor (2008a). However, unlike the 2007 convective season over the southern High Plains, downburst activity was more evenly distributed throughout the 2008 convective season.

Documentation of microbursts in the INL mesonet domain revealed that microburst activity occurred primarily during the afternoon. This preference for afternoon microburst activity underscores the importance of solar heating of the boundary layer in the process of convective downdraft generation over southeastern Idaho. The well-mixed moisture profile and relative humidity gradient that results from diurnal heating fostered a favorable environment for microbursts that would occur due to the evaporation of precipitation in the dry sub-cloud layer. Water vapor satellite imagery, RAOBs and GOES sounding profiles provided evidence that significant mid-level moisture promoted precipitation loading as an initiating mechanism for downbursts over the INL domain. The combination of precipitation loading and the presence of a relatively deep and dry convective boundary layer favored a microburst environment that was effectively captured by the GOES imager-derived microburst product. Thus, derivation of an algorithm that incorporates GOES-11 bands 3, 4, and 5 appears to be effective in indicating a favorable thermodynamic environment for microbursts over southeastern Idaho as well as other regions in the intermountain western U.S.

## 5. SUMMARY AND CONCLUSIONS

As documented in Pryor (2008a) and in this paper, and proven by statistical analysis, the GOES sounder MWPI product has demonstrated capability in the assessment of wind gust potential over the southern High Plains. In addition, a new multispectral GOES imager product has been developed and evaluated to assess downburst potential over the western United States. This microburst risk product image incorporates GOES-11 bands 3, 4, and 5 to sample the warm-season pre-convective environment and derive moisture stratification characteristics of the boundary layer that would be relevant in the analysis of microburst potential. Case studies and statistical analysis for downburst events that occurred over southeastern Idaho during the 2007 and 2008 convective seasons demonstrated the effectiveness of the imager product with a strong correlation between risk values and microburst wind gust magnitude. The GOES-11 imager microburst product has been found to be effective in indicating the potential for dry microbursts. However, as noted by Caracena and Flueck (1988), the majority of microburst days during JAWS were characterized by environments intermediate between the dry and wet extremes (i.e. hybrid). As noted in Pryor (2008a), the MWPI product is especially useful in the inference of the presence of intermediate or "hybrid" microburst environments, especially over the Great Plains region.

Future development effort will entail the implementation of a color enhancement to the GOES-11 imager product to highlight regions high microburst risk, in a similar manner to the product image displayed in Figure 2 and existing GOES-imager derived fog detection products. This product provides a higher spatial (4 km) and temporal (30 minutes) resolution than is currently offered by the GOES sounder microburst products and thus, should provide useful information to supplement the sounder products in the convective storm nowcasting process.

**Acknowledgements**

The author thanks Mr. Derek Arndt (Oklahoma Climatological Survey)/Oklahoma Mesonet, Dr. Kirk Clawson (NOAA/ARL/FRD), and Dr. John Schroeder (Texas Tech University)/West Texas

Mesonet for the surface weather observation data used in this research effort. The author also thanks Jaime Daniels (NESDIS) for providing GOES sounding retrievals displayed in this paper.